\newif\ifsingle
\singletrue 
 
\ifsingle
\documentclass[12pt,draftclsnofoot, onecolumn]{IEEEtran}		
\else		
\documentclass[10pt,final, twocolumn]{IEEEtran}
\fi

\usepackage{times}
\usepackage{amsmath,dsfont}
\usepackage{amssymb,amsthm}
\usepackage{epsfig,verbatim}
\usepackage{subfigure}
\usepackage{setspace}
\usepackage{color}
\usepackage{cite}
\usepackage{epstopdf}
\usepackage{graphics}
\usepackage{accents}
\usepackage{acronym}
\usepackage[bookmarks,colorlinks]{hyperref}
\usepackage{booktabs}
\usepackage{mathtools}
\usepackage{float}

\newtheorem{definition}{Definition}

\newcommand{\myVec}[1]{{\boldsymbol{#1}}}
\newcommand{\myMat}[1]{{\boldsymbol{#1}}}
\newcommand{\mySet}[1]{\mathcal{#1}}

\newcommand{\E}{\mathds{E}}		 			
\newcommand{\myX}{{\myVec{x}}}			 	
\newcommand{\myS}{{\myVec{s}}}			 	
\newcommand{\lenX}{n}			 			
\newcommand{\lenS}{k}			 			
\newcommand{\Quan}[2]{Q_{#1}^{#2}}			
\newcommand{\myI}{{\myMat{I}}}			 		
\newcommand{\Nantennas}{n_t}						
\newcommand{\Nusers}{n_u}							
\newcommand{\Ppilots}{P}							
\newcommand{\Tpilots}{\tau_p}						
\newcommand{\myH}{\myVec{h}}

\newcommand{\myY}{{\myVec{y}}}			 			
\newcommand{\myWV}{\myVec{w}}			 			
\newcommand{\myPhi}{\myMat{\Phi}}			 		
\newcommand{\Real}{{\rm Re}}
\newcommand{\Imag}{{\rm Im}}

\newcommand{\myTh}{{\myVec{s}}}			 		
\newcommand{\myZ}{{\myVec{z}}}			 	
\newcommand{\Rate}{R}
\newcommand{\Ratio}{\rho}
\newcommand{\AsymDist}{\eta}

\newcommand{\TilM}{\tilde{M}}

\newcommand{\Pdf}[1]{f_{ #1}}
\newcommand{\lenZ}{p}			 			

\newcommand{\Vecdim}[1]{} 

\newcommand{\NetMap}{\psi_{\myVec{\theta}}}

\ifsingle

\newcommand{\includefig}[1]{\includegraphics[width = 0.65\columnwidth]{#1} 	\vspace{-0.2cm}}
\else

\newcommand{\includefig}[1]{\includegraphics[width = 0.9\columnwidth]{#1} 	\vspace{-0.2cm}}

\fi 

\long\def\symbolfootnote[#1]#2{\begingroup\def\thefootnote{\fnsymbol{footnote}}\footnote[#1]{#2}\endgroup}

\acrodef{rv}[RV]{random variable}
\acrodef{mse}[MSE]{mean-squared error}
\acrodef{mmse}[MMSE]{minimum \ac{mse}}
\acrodef{bs}[BS]{base station}
\acrodef{dnn}[DNN]{deep neural network}
\acrodef{dft}[DFT]{discrete Fourier transform}
\acrodef{dt}[DT]{discrete-time}
\acrodef{ct}[CT]{continuous-time}
\acrodef{awgn}[AWGN]{additive white Gaussian noise}
\acrodef{wss}[WSS]{wide-sense stationary}
\acrodef{mimo}[MIMO]{multiple-input multiple-output}
\acrodef{map}[MAP]{maximum a-posteriori probability}
\acrodef{isi}[ISI]{intersymbol interference}
\acrodef{snr}[SNR]{signal-to-noise ratio}
\acrodef{pdf}[PDF]{probability density function}
\acrodef{lbg}[LBG]{Linde, Buzo, and Gray}
\acrodef{adc}[ADC]{analog-to-digital converter}
\acrodef{sgd}[SGD]{stochastic gradient descent}
\acrodef{dof}[DoF]{Degree-of-Freedom}
\acrodef{bpsk}[BPSK]{binary phase shift keying}
\acrodef{ber}[BER]{bit error rate}
\acrodef{csi}[CSI]{channel state information}
\acrodef{sbrnn}[SBRNN]{sliding bidirectional recursive neural network}

 \IEEEoverridecommandlockouts
\title{Deep Task-Based Quantization
}

\vspace{-0.2cm}

\author{
	\IEEEauthorblockN{Nir Shlezinger and Yonina C. Eldar\\
	} 
	\thanks{This project has received funding from the European Union’s Horizon 2020 research and innovation program under grant No. 646804-ERC-COG-BNYQ, and from the Israel Science Foundation under grant No. 0100101.}
	\thanks{Parts of this work were presented in the 2019 IEEE International Conference on Acoustics, Speech, and Signal Processing (ICASSP), Birghton, UK.}
	 \thanks{
		The authors are with the faculty of Mathematics and Computer Science, Weizmann Institute of Science, Rehovot, Israel (e-mail: nirshlezinger1@gmail.com;   yonina@weizmann.ac.il).
	}
	
	\vspace{-0.75cm}
	
}

\vspace{-0.5cm}

\begin{document}

\maketitle
\pagestyle{empty}
\thispagestyle{empty}
\begin{abstract}
	Quantizers play a critical role in digital signal processing systems.  
	Recent works have shown that the performance of quantization systems acquiring multiple analog signals using scalar  analog-to-digital converters (ADCs) can be significantly improved by properly processing the analog signals prior to quantization. However,  the design of such hybrid quantizers is quite complex, and their implementation requires  complete knowledge of the statistical model of the analog signal, which may not be available in practice. 
	In this work we design data-driven task-oriented quantization systems with scalar ADCs, which determine how to map an analog signal into its digital representation using deep learning tools. These representations are designed to facilitate the task of recovering underlying information from the quantized signals, which can be a set of parameters to estimate, or alternatively, a classification task.  By utilizing deep learning, we circumvent the need to explicitly recover the system model and to find the proper quantization rule for it. Our main target application is multiple-input multiple-output (MIMO) communication receivers, which simultaneously acquire a set of analog signals, and are commonly subject to constraints on the number of bits.  Our results indicate that, in a MIMO channel estimation setup, the proposed  deep task-bask quantizer is capable of approaching the optimal performance limits dictated by indirect rate-distortion theory, achievable using vector quantizers and requiring complete knowledge of the underlying statistical model. Furthermore, for a symbol detection scenario, it is demonstrated that the proposed approach can realize reliable bit-efficient hybrid MIMO receivers capable of setting their quantization rule in light of the task, e.g., to minimize the bit error rate.  	
\end{abstract}

\vspace{-0.2cm}
\section{Introduction}
\vspace{-0.1cm}
Digital signal processing systems operate on finite-bit representation of continuous-amplitude physical signals. The mapping of an analog signal into a digital representation  of a finite dictionary is referred to as {\em quantization} \cite{Gray:98}. 
This representation is commonly selected to accurately match the quantized signal, in the sense of minimizing a distortion measure, such that the signal can be recovered with minimal error from the quantized measurements  \cite{Berger:98}, \cite[Ch. 10]{Cover:06}. 
In many relevant scenarios, the task of the system is to recover some underlying parameters, and not to accurately represent the observed signal. 
In these cases, it was shown that by accounting for the system task in the design of the quantizers, namely, by utilizing {\em task-based quantization},  the  performance can be improved without increasing the number of bits used~\cite{Rodrigues:17,Shlezinger:18, Shlezinger:19, Salamatian:19b}.

In practice, quantizers are typically implemented using \acp{adc}, which operate on the input signal in a serial scalar manner. In such systems, the quantization rule is based on a uniform partition of a subspace of the real line, determined by the support of the quantizer. This quantization logic is very limited due to its simplicity: except for the specific case where the input is uniformly distributed over the support of the quantizer, uniform quantization is far from optimality \cite[Sec. 22]{Polyanskiy:15}, namely, a more accurate representation can be obtained with the same number of bits. Furthermore, such quantizers typically do not account for the system task, namely, they are {\em task-ignorant}. While the distortion induced by such inefficient quantization can be mitigated by assigning more bits for digital representation, i.e., using high-resolution quantizers, it can severely degrade the performance of bit-constrained systems.

Recent years have witnessed a growing interest in systems operating with low-resolution \acp{adc}. In particular, the power consumption of \acp{adc} typically grows with the bandwidth and the quantization resolution \cite{Walden:99}. To maintain feasible cost and power usage when acquiring multiple signals at large frequency bands,  low-resolution quantizers may be used. An example where such bit-constrained systems are popular is \ac{mimo} communication receivers, which simultaneously acquire and process multiple analog signals in order to recover the transmitted symbols and/or estimate the underlying channel, i.e., for a specific task. \ac{mimo} receivers operating at large spectral bands, e.g., millimeter wave systems \cite{Xiao:17}, are commonly designed to acquire the channel output with low-resolution quantizers, and a large body of work focuses on schemes for carrying out the aforementioned tasks from coarsely discretized measurements, see, e.g., \cite{Li:17,Mo:18,Mollen:17,Choi:16,Jacobsson:17,Pirzadeh:18,Studer:16}


Quantizers are inherently non-linear systems. Hence, the design and implementation of practical quantizers which provide an accurate discrete representation while accounting for the system task, is difficult in general. Two notable challenges are associated with designing such task-based quantization systems: 
$1)$ In order to design the quantization scheme, one must have full knowledge of the stochastic model of the underlying signal \cite{Berger:98,Gray:98}, which may be unavailable in practice;
$2)$ Even when the stochastic model is perfectly known, the scalar continuous-to-discrete rule which minimizes the representation error is generally unknown for most distributions under finite resolution quantization \cite[Ch. 23.1]{Polyanskiy:15}.
A possible approach to tackle the second challenge is to use a uniform quantization rule, while applying additional processing in analog prior to quantization, resulting in an analog-digital hybrid system \cite{Stein:19,Mo:17}. While such hybrid systems were shown to result in substantially improved performance for signal recovery tasks under bit constraints \cite{Shlezinger:18,Shlezinger:19,Salamatian:19b}, their design is commonly restricted to a subset of analog mappings, e.g., linear processing \cite{Shlezinger:18}; and specific stochastic models, such as Gaussian observations \cite{Shlezinger:19,Salamatian:19b}. Furthermore, these model-based quantization systems assume uniform quantizers, hence, they do not exploit the ability to utilize arbitrary quantization rules, while requiring accurate knowledge of the underlying statistical model.     

An alternative approach to inferring the quantization system from the model, is to learn it from a set of training samples in a data-driven fashion. In particular, by utilizing machine learning methods, one can  implement task-based quantizers without the need to explicitly know the underlying model and to analytically derive the proper quantization rule.  
Existing works on deep learning for quantization typically focus on image compression \cite{Toderici:17, Agustsson:17,Balle:16, Balle:18, Johnston:17},
where the goal is to represent the analog image using a single quantization rule, i.e., non task-based quantization. Alternatively, a large body of deep learning related works consider \ac{dnn} model compression \cite{Han:15,Ullrich:17,Hubara:18}, where a \ac{dnn}  operates with quantized instead of continuous weights. 
The work \cite{Yang:19} used \acp{dnn} to compress and quantize high-dimensional channel state information in a massive \ac{mimo} feedback setup.
The design of \acp{dnn} for processing one-bit quantized measurements in the digital domain, i.e., in the presence of task-ignorant quantizers, was considered for  signal recovery in \cite{Khobahi:18}; while \ac{dnn}-based \ac{mimo}  receivers with one-bit quantizers were studied in \cite{Balevi:19,Choi:18}. To the best of our knowledge, despite the importance of  quantization with scalar \acp{adc} in digital signal processing, the application of deep learning in such systems has not yet been studied.


In this paper we consider the design of data-driven  task-based quantizers, utilizing scalar \acp{adc}. Following \cite{Shlezinger:18,Shlezinger:19, Salamatian:19b}, we propose a hybrid quantization system in which the analog mapping, the quantization rule, and the digital processing, are learned from training in an end-to-end fashion. The operation of the scalar \acp{adc} is modeled as an intermediate activation layer. Unlike previous works which combined fixed uniform quantizers as part of a neural network \cite{Balle:16,Balle:18,Yang:19}, 
our method is specifically designed for learning scalar quantization mappings.
We consider two generic tasks: estimating a set of parameters taking values in a continuous set from the quantized observations, and classifying the acquired signals. Our main target application is bit constrained \ac{mimo} receivers, in which these tasks may represent, for example, channel estimation and symbol detection, respectively.

Since continuous-to-discrete mappings applied in the quantization process are inherently non-differentiable, standard deep learning training algorithms, such as \ac{sgd}, cannot be applied in a straight-forward manner. 
To overcome this difficulty, previous works used a  simplified model of the  quantizer, in which  the quantization error is replaced by additive i.i.d. noise \cite{Balle:16,Balle:18,Yang:19}. 
As the quantization error is a deterministic function of the analog input \cite{Widrow:96}, the resulting model is relatively inaccurate, inducing a mismatch which, as we numerically demonstrate, degrades the ability to properly optimize the system in light of the task. Furthermore, this model is  limited to fixed uniform continuous-to-discrete mappings, namely, the quantization mapping cannot be learned during training. 
Here, we approximate the continuous-to-discrete mapping with a differentiable one during training which faithfully represents the operation of the quantizer, facilitating the application of back-propagation, while allowing to learn the quantization mapping as part of an end-to-end network.  

We numerically evaluate the performance of our proposed \ac{dnn}-based system in \ac{mimo} communication scenarios. We first consider channel estimation, and compare our data-driven task-based quantizer to previous channel estimators from task-ignorant quantized measurements, as well as to the model-based task-based quantization system proposed in our previous work \cite{Shlezinger:18}. We also compare with the fundamental limits on channel estimation performance in \ac{mimo} systems with quantized observations, derived using indirect rate-distortion theory, which are achievable using optimal vector quantizers \cite[Ch. 23]{Polyanskiy:15}. 
Our results demonstrate that, even when the \ac{dnn}-based quantizer is trained with samples taken from setups with different \ac{snr}, it is still able to approach the performance of the optimal task-based quantizers with \acp{adc} for varying \acp{snr}, which is within a small gap of the fundamental performance limits. 

Next, we test the data-driven quantizer for the task of symbol detection in multi-user \ac{mimo} communications. Here, we show that our  quantizer achieves performance which is  comparable to applying the \ac{map} rule without any quantization constraints, and is notably more robust to inaccurate \ac{csi}.
Furthermore,   our deep task-based quantizer significantly outperforms the previously used approach of modeling quantization as additive noise during training, and we illustrate that the gap stems from the usage of a more accurate model for the quantization mapping.
We also discuss how the proposed approach can be exploited to construct trainable task-based \acp{adc}, by combining neuromorphic electronic systems \cite{Mead:90} with digital neural networks, giving rise to robust, efficient, and accurate, data-driven methods for  acquisition of analog signals.

The rest of this paper is organized as follows: 
Section~\ref{sec:Preliminaries} formulates the problem; 
Implementation of the data-driven task-based quantizer is presented in Section~\ref{sec:DNN}.   
Section~\ref{sec:Application} numerically evaluates the proposed quantizer in \ac{mimo} communication scenarios. 
Finally, Section~\ref{sec:Conclusions} provides some concluding remarks.

Throughout the paper, we use boldface lower-case letters for vectors, e.g., ${\myVec{x}}$,
and boldface upper-case letters for matrices,  e.g., $\myMat{M}$. 
Sets are denoted with calligraphic letters, e.g., $\mathcal{X}$. 
We use $\myI_{n}$ to represent the $n \times n$ identity matrix. 
Transpose,  Euclidean norm,  stochastic expectation,  real part, and imaginary part are written as  $(\cdot)^T$,  $\left\|\cdot\right\|$, $\E\{ \cdot \}$, $\Real\left(\cdot\right)$, and $\Imag\left(\cdot\right)$,   respectively, 
 $\mySet{R}$ is the set of real numbers, and $\mySet{C}$ is the set of complex numbers. 


\vspace{-0.2cm}
\section{Preliminaries and Problem Statement}
\label{sec:Preliminaries}
\vspace{-0.1cm}
\subsection{Preliminaries in Quantization Theory}
\label{subsec:Pre_Works}
\vspace{-0.1cm}
To formulate the problem, we first briefly review the standard quantization setup. 
While parts of this review also appear in our previous work \cite{Shlezinger:18}, it is included for completeness.  
We begin with the definition of a quantizer:
\begin{definition}[Quantizer]
	\label{def:Quantizer}
	A quantizer $\Quan{M}{n,k}\left(\cdot \right)$ with $\log M$ bits, input size $n$, input alphabet $\mySet{X}$, output size $k$, and output alphabet $\hat{\mySet{X}}$, consists of: 
	{\em 1)} An  encoding function $f_n: \mySet{X}^n \mapsto \{1,2,\ldots,M\} \triangleq \mySet{M}$ which maps the input into a discrete index.
	{\em 2)} A decoding function  $g_k: \mySet{M} \mapsto \hat{\mySet{X}}^k$ which maps each index $i \in \mySet{M}$ into a codeword $\myVec{q}_i \in  \hat{\mySet{X}}^k$. 
\end{definition}
We write the output of the quantizer with input $\myX\Vecdim{n} \in \mySet{X}^n$ as $\hat{\myX}\Vecdim{k} = g_k\left( f_n\left( \myX\Vecdim{n}\right) \right) \triangleq \Quan{M}{n,k}\left( \myX\Vecdim{n}\right)$. 
{\em Scalar quantizers} operate on a scalar input, i.e., $n=1$ and $\mySet{X}$ is a scalar space, while {\em vector quantizers} have a multivariate input. 
When the input size and the output size are equal,  $n=k$, we write $\Quan{M}{n}\left(\cdot \right) \triangleq \Quan{M}{n,n}\left(\cdot \right)$.

In the standard quantization problem, a $\Quan{M}{n}\left(\cdot \right)$ quantizer is designed to minimize some distortion measure  $d:\mySet{X}^n\times\hat{\mySet{X}}^n \mapsto \mySet{R}^+$  between its input and its output. 
The performance of a quantizer is  characterized using two measures: the quantization rate, defined as $\Rate \triangleq \frac{1}{n}\log M$, and the expected distortion $\E\{d\left(\myX\Vecdim{n}, \hat{\myX}\Vecdim{n} \right)\}$. For a fixed input size $n$ and codebook size $M$, the optimal quantizer is
\begin{equation}
\label{eqn:OptQuantizer}
\Quan{M}{n, {\rm opt}}\left(\cdot \right) = \mathop{\arg \min}\limits_{\Quan{M}{n}\left(\cdot \right)} \E \left\{d\left(\myX, \Quan{M}{n}\left( {\myX} \right)\right)   \right\}.
\end{equation}
Characterizing the optimal quantizer via \eqref{eqn:OptQuantizer} and its trade-off between distortion and quantization rate is in general a very difficult task. Optimal quantizers are thus typically studied assuming either high quantization rate, i.e., $\Rate \rightarrow \infty$, see, e.g., \cite{Li:99}, or asymptotically large inputs, namely, $n \rightarrow \infty$, commonly with i.i.d. inputs, via rate-distortion theory \cite[Ch. 10]{Cover:06}. 


In {\em task-based quantization}, the design objective of the quantizer is some task 
other than minimizing the distortion between its input and output. 
In the following, we focus on the generic task of acquiring a random vector  $\myTh \in \mySet{S}^\lenS \subseteq \mySet{R}^\lenS$ from a  statistically dependent random vector $\myX \in \mySet{R}^\lenX$.  The set $\mySet{S}$ represents the possible values of the unknown vector: It can be continuous, representing an estimation task; discrete, for classification tasks; or binary, for detection tasks.
This formulation accommodates a broad range of applications, including channel estimation and symbol detection, that are the common tasks considered in bit-constrained hybrid \ac{mimo} communications receivers \cite{Shlezinger:19}, which are the main target systems considered in this work.  

When quantizing for the task of estimation, under the objective of minimizing the \ac{mse} distortion, i.e., $d(\myVec{s}, \hat{\myVec{s}}) = \| \myVec{s} - \hat{\myVec{s}} \|^2 $, it was shown in \cite{Wolf:70} that the optimal quantizer applies vector quantization to the \ac{mmse} estimate of the desired vector $\myVec{s}$ from the observed vector $\myVec{x}$. While the optimal system utilizes vector quantization, the fact that such pre-quantization processing can improve the performance in estimation tasks was also demonstrated in \cite{Shlezinger:18}, which considered scalar quantizers. However, it was also shown in \cite{Shlezinger:18} and \cite{Salamatian:19b} that the pre-quantization processing which is optimal with vector quantizers, i.e., recovery of the \ac{mmse} estimate of $\myVec{s}$ from $\myVec{x}$, is no longer optimal when using scalar quantization, and that characterizing the optimal pre-quantization processing in such cases is very difficult in general. 
The fact that processing the observations in the analog domain is beneficial in task-based quantization motivates the hybrid system model which is the focus of the current work, and detailed in the following subsection. Due to the difficulty in analytically characterizing the optimal hybrid system, we consider a data-driven design, described in Section \ref{sec:DNN}.  
%


\vspace{-0.2cm}
\subsection{Problem Statement}
\label{subsec:Pre_Problem}
\vspace{-0.1cm}
As discussed in the introduction, practical digital signal processing systems typically obtain a digital representation of physical analog signals using  scalar \acp{adc}. 
Since in such systems, each continuous-amplitude sample is converted into a discrete representation using a single quantization rule, this operation can be modeled using {\em identical scalar quantizers}. In this work we study the implementation of task-based quantization systems with  scalar \acp{adc} in a data-driven fashion.

\begin{figure*}
	\centering
	\includegraphics[width=16cm]{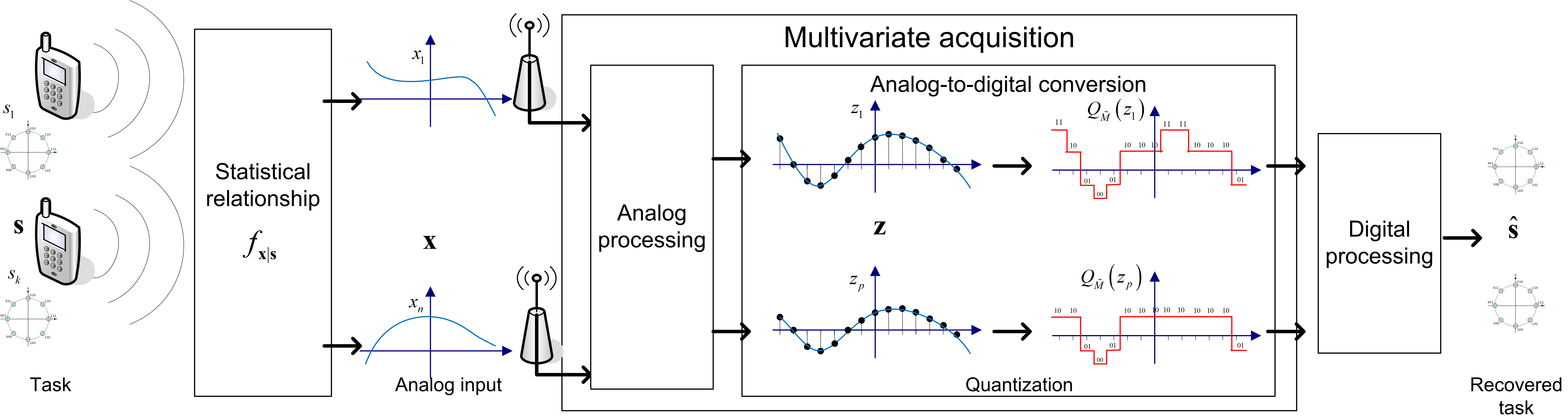}
	\vspace{-0.4cm}
	\caption{Hybrid task-based quantization system model. For illustration, the task is recovering a set of constellation symbols in uplink \ac{mimo} communications.}
\ifsingle
\else	
	\vspace{-0.4cm}
\fi
	\label{fig:SystemModel}
\end{figure*}

The considered signal acquisition system with scalar \acp{adc} is modeled using the hybrid setup depicted in Fig. \ref{fig:SystemModel}, where a set of analog signals are converted to digital in order to extract some desired information from them. This model can represent, e.g., sensor arrays or \ac{mimo} receivers, and specializes the case of a single analog input signal. While acquiring a set of analog signals in digital hardware includes both sampling, i.e., continuous-to-discrete time conversion, as well as quantization, namely, continuous-to-discrete amplitude mapping, we henceforth focus only the quantization aspect assuming a fixed sampling mechanism, and leave the data-driven design of the overall system for future investigation. 

 We consider the recovery of an unknown random  vector $\myS \in \mySet{S}^\lenS$ based on an observed vector $\myX \in \mySet{R}^\lenX$ quantized with up to $\log M$ bits. The observed $\myX$ is related to $\myS$ via a conditional probability measure $\Pdf{\myX|\myS}$, which is assumed to be unknown. For example, in a communications setup. the conditional probability measure $\Pdf{\myX|\myS}$ encapsulates the noisy channel.  
 The input to the \ac{adc}, denoted $\myZ \in \mySet{R}^\lenZ$, where $\lenZ$ denotes the number of scalar quantizers, is obtained from $\myX$ using some pre-quantization mapping carried out in the analog domain. Then, $\myZ$ is quantized using an \ac{adc} modeled as $\lenZ$ identical scalar quantizers with resolution $\TilM \triangleq \lfloor M^{1/\lenZ}\rfloor$. The overall number of bits is $\lenZ \cdot \log \TilM \le \log M$. The \ac{adc} output is processed in the digital domain to obtain the quantized representation $\hat{\myS} \in \mySet{S}^\lenS$. 

Our goal is to design a generic machine-learning based architecture for task-based quantization with scalar \acp{adc}. The proposed system operates in a data-driven manner, namely, it is capable of learning the analog transformation, quantization mapping, and digital processing, from a training data set, consisting of $t$ independent realizations of $\myVec{s}$ and $\myVec{x}$, denoted $\{ \myVec{s}^{(i)}, \myVec{x}^{(i)}\}_{i=1}^{t}$.
In general, the training samples may be taken from a set of joint distributions, and not only from the true (unknown) joint distribution of $\myVec{s}$ and $\myVec{x}$, as we consider in our numerical study in Section~\ref{sec:Application}.
 We focus on two tasks which are relevant for \ac{mimo} receivers: An estimation task, in which $\mySet{S} = \mySet{R}$, representing, e.g., channel estimation; and classification, where $\mySet{S}$ is a finite set, modeling, e.g., symbol detection. Our design is based on machine-learning methods, and specifically, on the application of \acp{dnn}. 

\vspace{-0.2cm}
\section{Deep Task-Based Quantization} 
\label{sec:DNN}
\vspace{-0.1cm}
In the following, we present a deep task-based quantizer, which implements the system depicted in Fig. \ref{fig:SystemModel} in a data-driven fashion using \acp{dnn}. To that aim, we first discuss the proposed network architecture in Subsection \ref{subsec:DNNArch}. Then, in Subsection \ref{subsec:DNNTrain} we elaborate on the discrete-to-continuous mapping and its training method, and provide a discussion on the resulting system in Subsection \ref{subsec:DNNDiscussion}.

\vspace{-0.2cm}
\subsection{\ac{dnn} Architecture}
\label{subsec:DNNArch}
\begin{figure}
	\centering
	\includefig{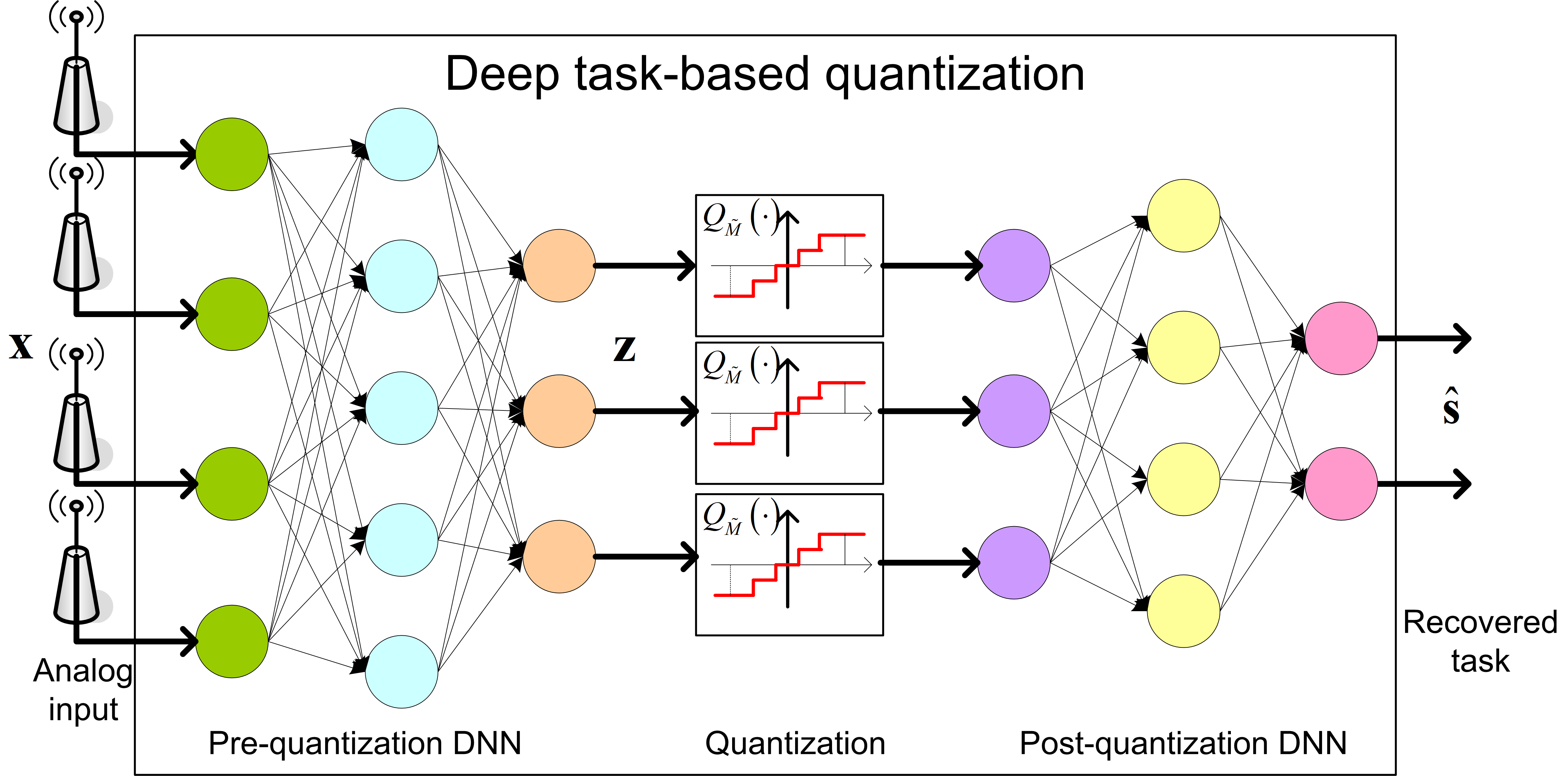}
	\caption{Deep task-based quantization system architecture.}
	\label{fig:DNNModel}
\end{figure}

\vspace{-0.1cm}
We propose to implement a data-driven task-based quantizer using machine-learning methods. In particular, we realize  the pre and post quantization mappings using dedicated \acp{dnn}, jointly trained in an end-to-end manner, as illustrated in Fig.~\ref{fig:DNNModel}.  

In the proposed architecture, the serial scalar \ac{adc}, which implements the continuous-to-discrete mapping, is modeled as an activation function between the two intermediate layers. The trainable parameters of this activation function determine the quantization rule, allowing it to be learned during training.
The \ac{dnn} structure cannot contain any skip connections between the multiple layers prior to quantization (analog domain) and those after quantization (digital domain), representing the fact that all analog values must be first quantized before processed in digital. The pre and post quantization networks are henceforth referred to as the {\em analog \ac{dnn}} and the {\em digital \ac{dnn}}, respectively. 
The system input is the $n \times 1$ observed vector $\myVec{x}$, and we use $\myVec{\theta}$ to denote the hyperparameters of the network. As detailed in Subsection \ref{subsec:Pre_Problem}, we consider two main types of tasks:
\begin{itemize}
	\item {\bf Estimation}: Here, the deep task-based quantizer should learn to recover a set of $\lenS$ unknown parameters taking values on a continuous set, i.e., $\mySet{S} = \mySet{R}$. By letting  $\NetMap(\cdot)$ denote the mapping implemented by the overall system, the output is  given by the $\lenS \times 1$ vector $\hat{\myVec{s}} = \NetMap(\myVec{x})$, which is used as a representation of the desired vector $\myVec{s}$. The loss function is the empirical \ac{mse}, given by
	\vspace{-0.1cm}
	\begin{equation}
	\label{eqn:LossFunc}
	\mathcal{L}(\myVec{\theta})  =\frac{1}{t}\sum_{j=1}^{t}\left\Vert \myS^{\left(j\right)}-\NetMap\Big(\myVec{x}^{(j)}\Big)\right\Vert _{2}^{2}.
	\vspace{-0.1cm}
	\end{equation} 
	\item {\bf Classification}: In such tasks, the deep task-based quantization should decide between a finite number of options based on its analog input. Here, $\mySet{S}$ is a finite set, and we use $|\mySet{S}|$ to denote its cardinality. The last layer of the digital \ac{dnn} is a softmax layer, and thus the network mapping  $\NetMap(\cdot)$ is a $|\mySet{S}|^\lenS \times 1$ vector, whose entries represent the conditional probability for each different value of $\myVec{s}$ given the input $\myVec{x}$. By letting $\NetMap(\myVec{x} ; \myVec{\alpha})$ be the output value corresponding to $\myVec{\alpha} \in \mySet{S}^{\lenS}$, the decision is selected as the most probable one, i.e., $\hat{\myVec{s}} = \arg\max_{\myVec{\alpha} \in \mySet{S}^{\lenS}}  \NetMap(\myVec{x} ; \myVec{\alpha})$.  	 The loss function is the empirical cross-entropy, given by
	\vspace{-0.1cm}
	\begin{equation}
	\label{eqn:LossFuncCE}
	\mathcal{L}(\myVec{\theta})  =\frac{1}{t}\sum_{j=1}^{t} -\log  \NetMap\Big(\myVec{x}^{(j)} ; \myVec{s}^{(j)} \Big).
	\vspace{-0.1cm}
	\end{equation} 
\end{itemize} 
 
 By utilizing \acp{dnn}, we expect the resulting system to be able to approach the optimal achievable distortion for fixed quantization rate $\Rate = \frac{1}{\lenX} \log M$ and input size $n$, without requiring explicit knowledge of the underlying distribution $\Pdf{\myVec{x}|\myVec{s}}$. Such performance is illustrated in the numerical example presented in Subsection \ref{subsec:App_Sim}.
 
The proposed architecture is generic, and its main novelty is in the introduction of the learned quantization layer, detailed in the following subsection. Our structure can thus be combined with existing dedicated networks, which are trainable in an end-to-end manner, as a form of transfer learning. For example, \acp{sbrnn} were shown to achieve good performance for the task of symbol detection in non-quantized communication systems with long memory \cite{Liao:19}. Consequently, one can design a deep symbol detector operating under quantization constraints, as common in, e.g., millimeter wave communications \cite{Xiao:17}, by implementing the digital \ac{dnn} of Fig. \ref{fig:DNNModel} as an \ac{sbrnn}. In this work we focus on fully-connected analog and digital \acp{dnn}, and leave the analysis of combination with dedicated networks to future investigation.

 
 
\vspace{-0.2cm}
\subsection{Quantization Activation}
\label{subsec:DNNTrain}
\vspace{-0.1cm}
Our proposed deep task-based quantizer implements scalar quantization as an intermediate activation in a joint analog-digital hybrid \ac{dnn}. This layer converts its continuous-amplitude input into a discrete digital representation.  The non-differentiable nature of such continuous-to-discrete mappings induces a major challenge in applying \ac{sgd} for optimizing the hyperparameters of the network.  
In particular, quantization activation, which can be modeled as a superposition of step functions determining the continuous regions jointly mapped into a single value, nullifies the gradient of the cost function. Consequently,  straight-forward application of \ac{sgd} fails to properly set the pre-quantization network.
To overcome this drawback, we first review the common approach, referred to henceforth as {\em passing gradient}, after which we propose a new method, referred to as {\em soft-to-hard quantization}.

\subsubsection{Passing Gradient}
\label{sub:PassGradient}
In this approach   the quantized values are modeled as the analog values corrupted by mutually independent i.i.d. noise  \cite{Balle:16, Balle:18, Yang:19}, and thus quantization does not affect the back-propagation procedure. Since the quantization error is deterministically determined by the analog value \cite{Widrow:96}, the resulting model is quite inaccurate. Specifically, while under some input distributions, the quantization noise can be modeled as being {\em uncorrelated} with the input \cite{Widrow:96}, they are not mutually independent. In fact, in order for the quantization error to be independent of the input, one should use substractive dithered quantization \cite{Wannamaker:00}, which does not represent the operation of practical \acp{adc}. Consequently, using this model for quantization during training results in a mismatch between the trained system and the tested one.  

Under this model, the continuous-to-discrete mapping is fixed, representing, e.g., uniform quantization, and the training algorithm back-propagates the gradient value intact through the quantization layer. 
 An illustration of this approach is depicted in Fig. \ref{fig:Passing_Gradient}. 
We expect the resulting system to obtain poor performance  when non-negligible distortion is induced by the quantizers. In our numerical study presented in Subsection \ref{subsec:App_SimDet}, it is illustrated that this method   achieves relatively poor performance at low quantization rates, where scalar quantization induces an error term which is non-negligible and depends on the analog input. It is therefore desirable to formulate a network structure which  accounts for the presence of  scalar quantizers during training, and is not restricted to  fixed  uniform quantizers.

\subsubsection{Soft-to-Hard Quantization}
\label{sub:soft_hard_quant}
\begin{figure}
	\centering
	\subfigure[Passing gradient quantization system]{\label{fig:Passing_Gradient}\includegraphics[scale=0.5]{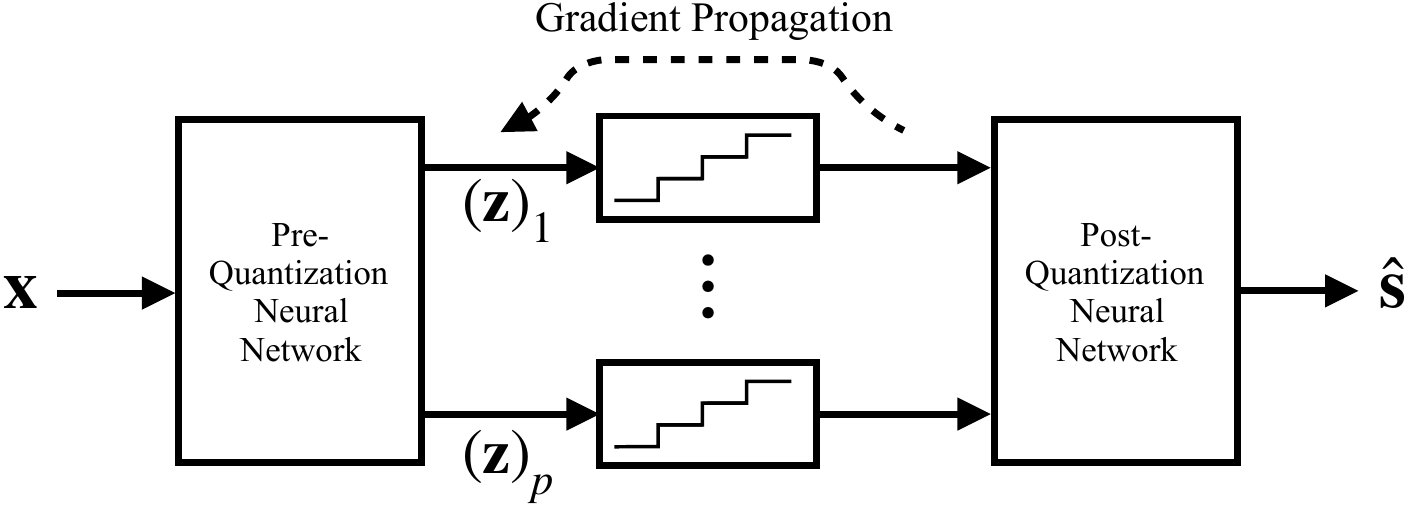}}
	\subfigure[Soft-to-hard quantization system]{\label{fig:Tanh_Quantization}\includegraphics[scale=0.5]{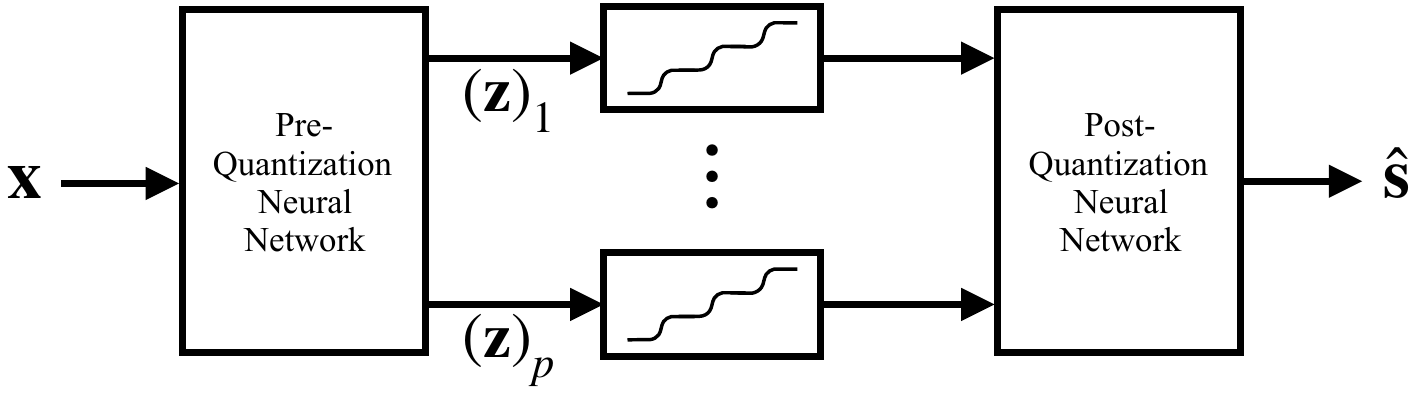}}
	\caption{Task-based deep quantization architectures.}
	\label{fig:DnnArchs}
\end{figure}

Our proposed approach  is based on approximating the non-differentiable quantization mapping by a differentiable one. Here, we replace the continuous-to-discrete transformation with a non-linear activation function which has approximately the same behavior as the quantizer, as illustrated in Fig. \ref{fig:Tanh_Quantization}. Specifically, we use a sum of shifted hyperbolic tangents, which are known to closely resemble step functions in the presence of large magnitude inputs. The resulting scalar quantization mapping is given by:
	\vspace{-0.1cm}
\begin{equation}
\label{eqn:tanh}
\tilde{q}_{\tilde{M}}(x)=\sum_{i=1}^{\tilde{M}-1}a_{i}\tanh\left(c_{i}\cdot   x-b_{i} \right),
	\vspace{-0.1cm}
\end{equation}
 where $\{a_i, b_i, c_i \}$ are a set of real-valued parameters. Note that as the parameters $\{c_i\}$ increase, the corresponding hyperbolic tangents approach step functions. Since we use a differentiable activation to approximate a set of non-differentiable functions  \cite{Agustsson:17}, we refer to this method as {\em soft-to-hard quantization}. 

In addition to learning the weights of the analog and digital \acp{dnn}, this soft-to-hard approach allows the network to learn its quantization activation function, and particularly, the best suitable constants $\{a_{i}\}$ (the amplitudes) and $\{b_{i}\}$ (the shifts). These tunable parameters are later used to determine the decision regions of the scalar quantizer, resulting in a learned quantization mapping.  
The parameters $\{c_i\}$, which essentially control the resemblance of \eqref{eqn:tanh} to an actual continuous-to-discrete mapping, do not reflect on the quantization decision regions (controlled by $\{b_{i}\}$) and their associated digital values (determined by $\{a_{i}\}$), and are thus not learned from training. The  set $\{c_i\}$  can be either set according to the quantization resolution $\TilM$, or alternatively, modified  using annealing-based optimization \cite{Rose:92}, where $\{c_i\}$ are manually increased during training.
The proposed optimization is achieved by including the parameters  $\{a_i, b_{i}\}$  as part of the network hyperparameters $\myVec{\theta}$. 
Due to the differentiability of  \eqref{eqn:tanh}, one can now apply standard \ac{sgd} to optimize the overall network, including the analog and digital \acp{dnn} as well as the quantization rule, in an end-to-end manner.

\begin{figure}
	\centering
	\includefig{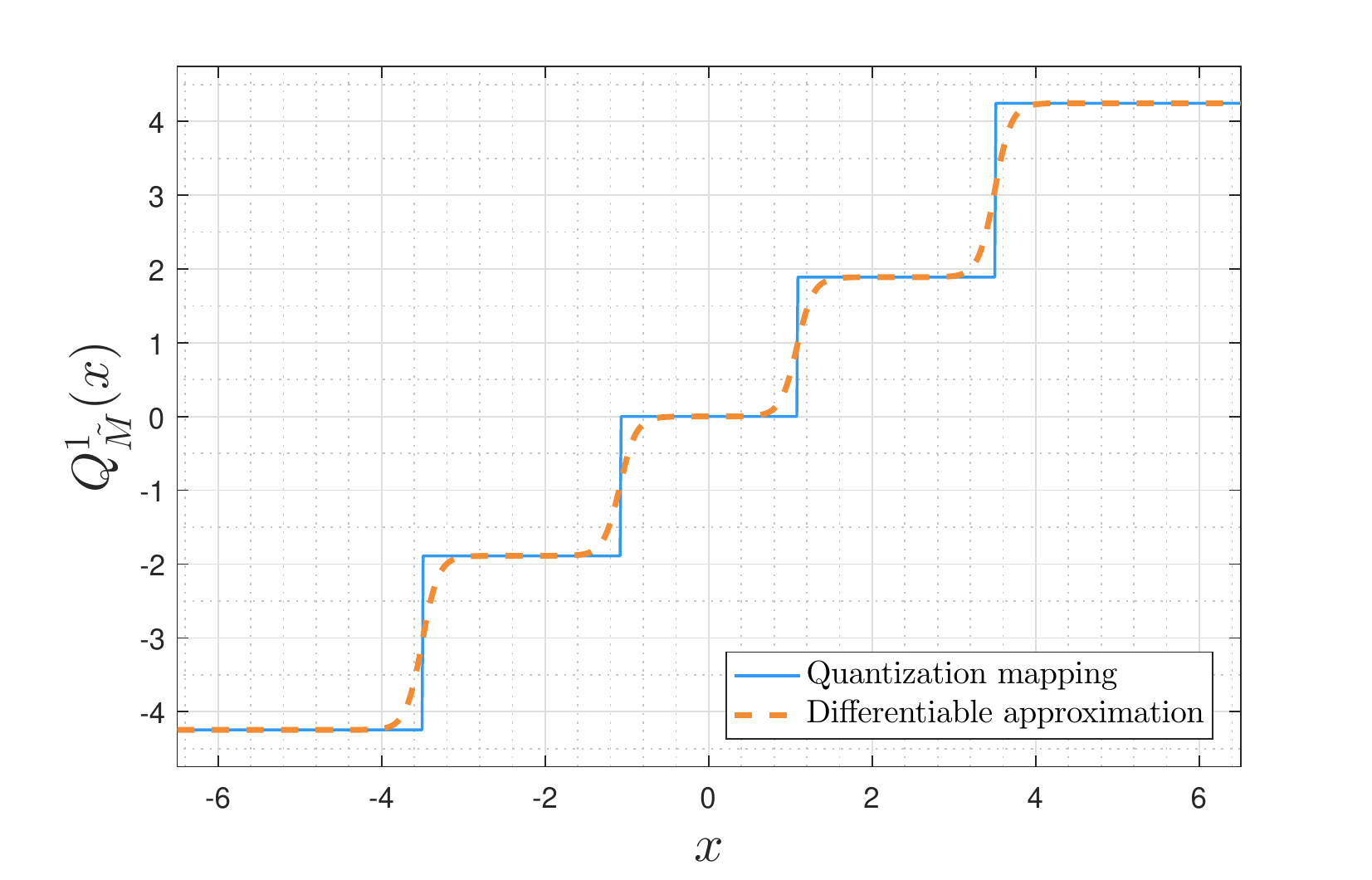} 
	\caption{Soft-to-hard quantization rule illustration.}
	\label{fig:Tanh_Aproximation}
\end{figure}

Once training is concluded,  we replace the learned $\tilde{q}_{\tilde{M}}(\cdot)$ activation \eqref{eqn:tanh} with a scalar quantizer whose decision regions are dictated by the tunable parameters $\{a_i, b_i \}$. 
In particular, since $\tanh(c \cdot x - b) = 0$ for $x = \frac{b}{c}$, we use the set $\left\{ \frac{b_{i}}{c_{i}}\right\}$ to determine the decision regions of the quantizer, and set the value of $\tilde{q}_{\tilde{M}}(x)$ at each decision region center as its corresponding representation level. 
Without loss of generality, we assume that $\frac{b_{0}}{c_{0}}\leq\frac{b_{1}}{c_{1}}\leq\ldots\leq\frac{b_{\tilde{M}-1}}{c_{\tilde{M}-1}}$ (when this conditions is not satisfied, the parameters are sorted and re-indexed accordingly).  The resulting quantizer is given by
	\vspace{-0.1cm}
\begin{equation}
Q_{\tilde{M}}^{1}(x)=\begin{cases}
-\sum_{i=1}^{\tilde{M}-1}a_{i} & x\leq \frac{b_{0}}{c_{0}}\\
\tilde{q}_{\tilde{M}}\left({\frac{b_{i}}{2c_{i}}+\frac{b_{i+1}}{2c_{i+1}}}\right) & \frac{b_{i}}{c_{i}}<x\leq \frac{b_{i+1}}{c_{i+1}}\\
\sum_{i=1}^{\tilde{M}-1}a_{i} & \frac{b_{\tilde{M}-1}}{c_{\tilde{M}-1}}<x.
\end{cases}
\label{eqn:quantizedTanh}
\end{equation} 
An illustration of how the differentiable mapping \eqref{eqn:tanh} is converted into a continuous-to-discrete quantization rule via \eqref{eqn:quantizedTanh} is depicted in Fig. \ref{fig:Tanh_Aproximation}. The dashed smooth curve in Fig. \ref{fig:Tanh_Aproximation} represents the differentiable function after training is concluded, and the straight curve is the resulting scalar quantizer.

In the simulations study presented in Subsection \ref{subsec:App_Sim}, it is illustrated that the proposed method, which faithfully represents the presence of scalar quantizers during training and is capable of optimizing their decision regions, can outperform  the model-based \ac{mse} minimizing task-based quantizer with scalar \acp{adc} of \cite{Shlezinger:18}, which requires complete knowledge of the underlying model, yet is restricted to uniform quantizers.

%
%
%
%
%
%

\vspace{-0.2cm}
\subsection{Discussion}
\label{subsec:DNNDiscussion}
\vspace{-0.1cm}
The deep task-based quantizer proposed in Subsection \ref{subsec:DNNArch} implements hybrid multivariate  acquisition using a set of identical scalar \acp{adc} with learned decision regions, combined with \ac{dnn}-based analog and digital transformations. While realizing \acp{dnn} in digital can be done in software, analog \acp{dnn} requires dedicated tunable hardware weights and activations. 
Such hardware networks, commonly referred to as neuromorphic electronic systems \cite{Mead:90}, implement configurable \acp{dnn} as analog components. Recent advances in memristors technology substantially facilitate the implementation of these hardware devices \cite{Danial:18}, contributing to the feasibility of our proposed deep task-based quantizer.

It is noted that in some applications, constrained analog structures may be preferable. For example, in \ac{mimo} receivers with a large number of antennas, i.e., massive \ac{mimo}, pre-quantization analog processing is commonly limited to phase shifting \cite{Stein:19}. In this case, the analog \ac{dnn} is replaced with a single layer whose weights are restricted to have a unit magnitude, and this constraint has to be accounted for in training. Here we focus on generic analog \acp{dnn}, in which the weights are not constrained.  

Our task-based quantizer can thus be implemented as a system consisting of adjustable analog hardware, configurable scalar quantizers, and software. The natural approach to set the hyperparameters of the network would be to train the system model offline in software using an a-priori acquired training set. The network weights and quantization decision regions obtained from this trained model can be then configured into the hardware components and the tunable \acp{adc}, resulting in the desired task-based quantization system.

One can also envision an online trainable task-based quantizer, which is capable of further tuning its hyperparameters in real-time to track dynamic environments, as in, e.g., \cite{Shlezinger:19a}. For example, a communication receiver using a deep task-based quantizer for symbol detection, can exploit a-priori knowledge of pilot sequences as labels corresponding to inputs acquired in real-time. A major challenge in implementing such a system stems from the fact that both the labels $\{\myVec{s}^{(j)}\}$ as well as the inputs $\{\myVec{x}^{(j)}\}$ are required in order to update a network coefficients using conventional training algorithms, e.g., \ac{sgd}. However, in our system the digital processor does not have  direct access to the analog signal, but only to its quantized digital representation. 
Consequently, if the processor  only utilizes digital values, it can only train the digital \ac{dnn} using \ac{sgd}. 
This challenge may be handled by allowing access to a high resolution quantized version of the analog signals,  acquired in the specific time instances for which labels are available. An alternative approach is to utilize an error-correction based update algorithm \cite{Widrow:90} instead of \ac{sgd}, or reinforcement learning methods \cite{Sutton:18}, since these techniques typically do not require direct access to the network input. 

\vspace{-0.2cm}
\section{Application to MIMO Receivers}
\label{sec:Application}
\vspace{-0.1cm}
While the generic deep task-based quantizer proposed in Section \ref{sec:DNN} is applicable to a multitude of different setups, our main target application, studied in this section, is uplink multi-user \ac{mimo} communications. The problem of  \ac{mimo} communications with low-resolution quantization is the focus of many recent works, including, e.g., \cite{Li:17,Choi:16,Mo:17,Shlezinger:19,Balevi:19}. Here, we consider  a single cell  multi-user \ac{mimo} system, in which $\Nusers$ single antenna users are served by a \ac{bs} with $\Nantennas$ antennas, which operates under quantization constraints. We focus on two tasks encountered in such setups: The first is channel estimation detailed in Subsection \ref{subsec:App_Sim}, for which we are  capable of quantifying the performance gap of our system from optimality as well as comparing it to model-based designs. Then, in Subsection \ref{subsec:App_SimDet} we focus on symbol detection, which we treat as a classification task.


\vspace{-0.2cm}
\subsection{Channel Estimation Task}
\label{subsec:App_Sim}
\vspace{-0.1cm}
We first consider channel recovery, which is an estimation task commonly encountered in \ac{mimo} systems. We focus on a specific scenario for which we can compute both the fundamental performance limits, namely, a lower bound on the achievable recovery accuracy which holds for any bit constrained system, as well as the performance of the best hybrid system restricted to using linear operations and uniform quantization, derived in \cite{Shlezinger:18}. These performance measures, which correspond to model-based systems, are used as a basis for comparison to evaluate our proposed data-driven task-based quantizer. The main motivation for the study detailed in this subsection is thus to compare the performance achievable using our proposed deep task-based quantizer to model-based techniques and the fundamental performance limits in a specific scenario where these values are computable.

In the following, we consider a channel estimation task carried out in a time diversity duplexing manner as in \cite{Shlezinger:19}, using orthogonal pilot sequences of length $\Tpilots \ge \Nusers$.
We use  $\myPhi \in \mySet{C}^{\Tpilots \times \Nusers}$ to denote the known pilot sequence matrix, where the orthogonality of the pilots implies that $\myPhi^H \myPhi = \Tpilots \cdot \myI_{\Nusers}$, and $\Ppilots$ is the \ac{snr}.  
Additionally, let $\myH\in \mySet{C}^{\Nusers \cdot \Nantennas}$ be a random vector whose entires are  i.i.d. zero-mean unit-variance complex normal channel coefficients, and $\myWV \in \mySet{C}^{\Tpilots \cdot \Nantennas}$ be a random vector with i.i.d. zero-mean unit-variance complex normal entries mutually independent of $\myH$, representing the additive noise at the \ac{bs}. The observed signal $\myY \in \mySet{C}^{\Tpilots \cdot \Nantennas}$, used by the \ac{bs}  to estimate $\myH$, can be written as \cite[Eq. (4)]{Li:17}:
\begin{equation}
\myY = \sqrt{\Ppilots}\left(\myPhi \otimes \myI_{\Nantennas} \right) \myH + \myWV,
\label{eqn:MassiveMIMORel}
\end{equation}
where $\otimes$ is the Kronecker product. 

To put the setup in \eqref{eqn:MassiveMIMORel} in the framework of our problem formulation, which considers real-valued signals, we write the observations as $\myX = \big[ \Real\left(\myY \right)^T ,\Imag\left(\myY \right)^T \big]^T$ and the unknown channel as $\myS = \big[ \Real\left(\myH \right)^T ,\Imag\left(\myH \right)^T \big]^T$. Consequently, the number of measurements is $\lenX = 2 \cdot \Tpilots \cdot \Nantennas$, the number of unknown parameters is $\lenS = 2 \cdot \Nusers \cdot \Nantennas$, and their ratio is $\Ratio = \frac{\Tpilots}{\Nusers}$, which is not smaller than one. 

The performance measure for evaluating the quantization systems here is the average \ac{mse}, namely, $\eta = \frac{1}{\lenS}	\E \left\{\left\|\myS \! - \! \hat{\myS} \right\|^2   \right\}$. For the above model,    the average \ac{mmse}, which is  the optimal performance achievable  with no quantization constraints, is given by  $\tilde{\AsymDist} = \frac{1}{2(1 + \Ppilots\cdot \Tpilots)}$. In the presence of quantization constraints,  the optimal approach  is to quantize the \ac{mmse} estimate \cite{Wolf:70}, and the resulting average distortion  is obtained from  rate-distortion theory \cite[Ch. 10.3]{Cover:06} as 
\vspace{-0.1cm}
\begin{equation}
\AsymDist_{\rm opt} = \tilde{\AsymDist} + \frac{\Ppilots\cdot \Tpilots}{2(1 + \Ppilots\cdot \Tpilots)} 2^{-2\Ratio \cdot \Rate}.
\label{eqn:MassiveOpt}
\vspace{-0.1cm}
\end{equation} 
Note that $\AsymDist_{\rm opt}$ is achievable using optimal vector quantization in the limit  $\Nantennas \rightarrow \infty$. For finite $\Nantennas$ and scalar quantizers, \eqref{eqn:MassiveOpt} serves as a lower bound on the achievable performance. We thus refer to $\AsymDist_{\rm opt}$ as the {\em fundamental performance limit}.

We now numerically evaluate our proposed deep task-based quantizer, compared to the fundamental performance limit in \eqref{eqn:MassiveOpt}, as well as to the performance of the task-based quantizer with scalar uniform \acp{adc} designed in \cite{Shlezinger:18}, denoted $\eta_{\rm sc}$. 
It is noted that  while our proposed system can modify the quantization regions, the model of \cite{Shlezinger:18} assumes fixed uniform quantizers. Consequently, the  average \ac{mse} of the system of  \cite{Shlezinger:18}  does not necessarily lower bound the performance of our proposed system. We also note that the system of \cite{Shlezinger:18} requires full knowledge of the underlying statistical model, namely, the \ac{snr} as well as the distribution of $\myH$ and $\myWV$.

We simulate a multi-user \ac{mimo} network in which a BS equipped with $\Nantennas = 10$ antennas serves $\Nusers = 4$ users. 
We set the SNR to be $\Ppilots = 4$ and the number of pilots to  $\Tpilots = 12$. As in \cite{Li:17}, we fix the pilots matrix $\myPhi$  to be the first $\Nusers$ columns of the $\Tpilots \times \Tpilots$  discrete Fourier transform  matrix. 
In the implementation of the deep quantizers, we set the pre and post quantization \acp{dnn} to consist of linear layers. The motivation for using linear layers stems from the fact that for the considered setup, the \ac{mmse} estimate is a linear function of the observations. Furthermore, this setting guarantees fair comparison with the model-based system of \cite{Shlezinger:18}, which focused on linear analog and digital processing. 
Following \cite[Cor. 1]{Shlezinger:18}, we evaluate the average \ac{mse} of our proposed systems with $\lenZ = \lenS$ quantizers. 
We consider two training sets, both of size $t = 2^{15}$: In the first training set, representing {\em optimal training}, the realizations $\{ \myVec{s}^{(i)}, \myVec{x}^{(i)}\}_{i=1}^{t}$ are sampled from the true joint distribution of $\myS, \myX$; In the second training set, representing {\em \ac{snr} uncertainty}, $\{ \myVec{s}^{(i)}, \myVec{x}^{(i)}\}_{i=1}^{t}$ are sampled from the  joint distribution of $\myS, \myX$ with different values of $\Ppilots$, uniformly randomized over the set $[1,10]$ for each realization. 
At the end of the training session, we fix the quantizer to implement the continuous-to-discrete rule in \eqref{eqn:quantizedTanh}. 
We numerically evaluate our trained proposed deep quantizer using $2^{10}$ independent channel realizations.

\begin{figure}
	\centering
	\includefig{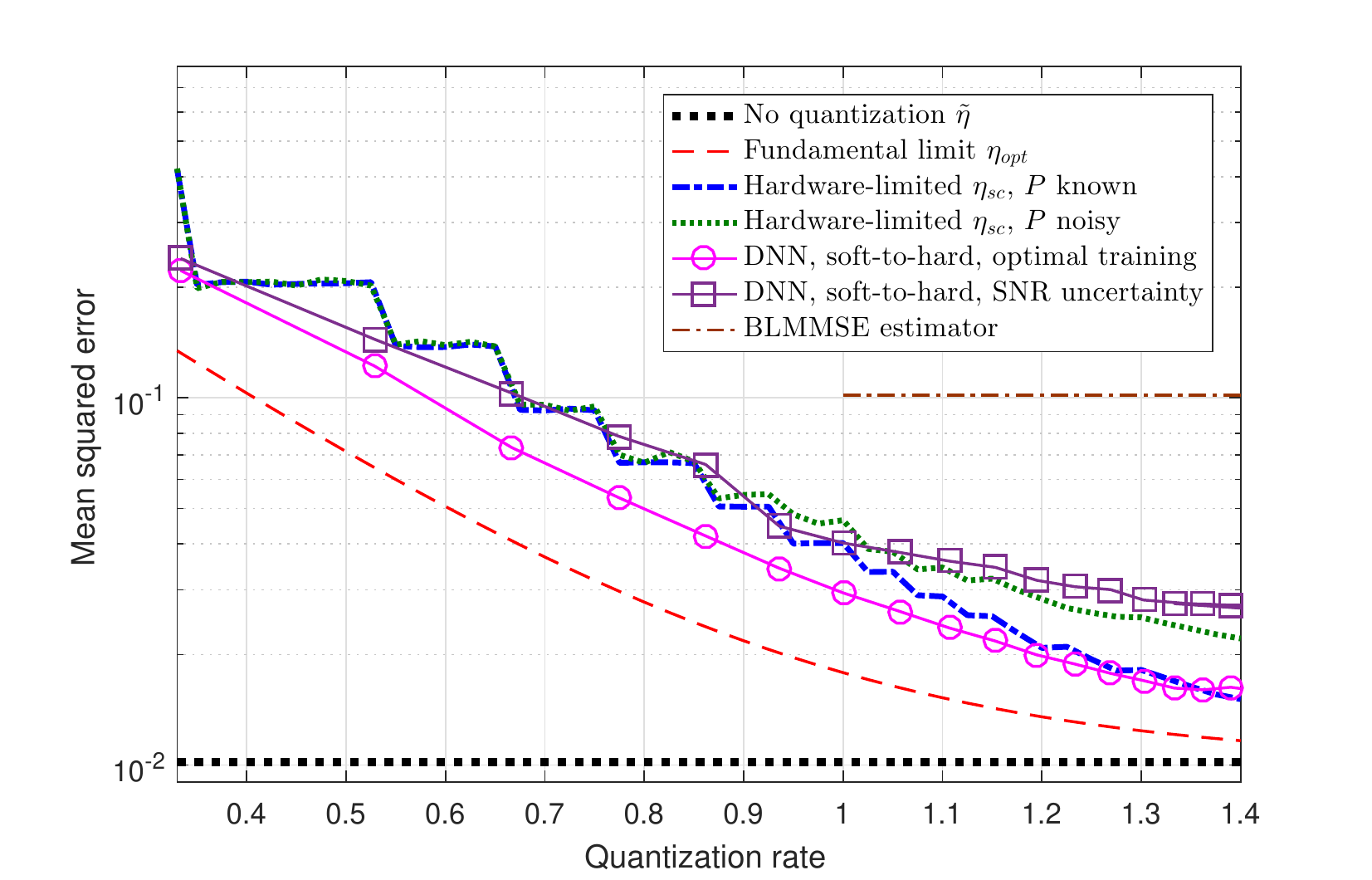} 
	\caption{Numerical \ac{mse} versus theoretical measures.}
	\label{fig:Numerical_Results}
\end{figure}

In Fig. \ref{fig:Numerical_Results} we depict the resulting performance versus the quantization rate $R = \frac{1}{n}\log M$ in the range $R \in [0.33,1.4]$. The empirical performance is compared to three theoretical measures: the \ac{mmse} $\tilde{\AsymDist}$; the fundamental performance limits of channel estimation from quantized measurements, given by $\AsymDist_{\rm opt}$ in \eqref{eqn:MassiveOpt}; and the performance of the analytically derived task-based quantizer with scalar \acp{adc} \cite{Shlezinger:18}, denoted  $\eta_{\rm sc}$. Since \cite{Shlezinger:18} requires perfect knowledge of the underlying model, and particularly of the \ac{snr} $\Ppilots$, and as this information may not be available accurately in practice, we also consider the case where \cite{Shlezinger:18} utilizes an estimation of $\Ppilots$ corrupted by zero-mean Gaussian noise with variance $\frac{\Ppilots}{4}$.   
Finally, we compute the average \ac{mse} of the BLMMSE estimator proposed in \cite{Li:17} via \cite[Eq. (15)]{Li:17}. Since the BLMMSE estimator quantizes the observed signal without analog pre-processing, it is applicable only for $R \ge 1$. 

Observing Fig. \ref{fig:Numerical_Results}, we note that the performance of our soft-to-hard deep quantizer is within a relatively small gap of the fundamental performance limits. 
Furthermore, the fact that the soft-to-hard method is not restricted to uniform quantizers allows it to outperform the model-based $\eta_{\rm sc}$, especially in lower quantization rates.   
Finally, we note that in the presence of \ac{snr} uncertainty, the performance of the soft-to-hard method is similar to $\eta_{\rm sc}$ with noisy \ac{snr} estimate, and that both outperform the BLMMSE estimator of \cite{Li:17}. This indicates that our proposed scheme is applicable also when the training data is not generated from the exact same distribution as the test data. 
Our results demonstrate the ability of deep task-based quantization to implement a feasible and optimal-approaching quantization system in a data-driven fashion. 

\vspace{-0.2cm}
\subsection{Symbol Detection Task}
\label{subsec:App_SimDet}
\vspace{-0.1cm}
The main task of a communication receiver is to recover the transmitted messages. Channel estimation, studied in the previous subsection, is intended to facilitate the recovery of the unknown symbols. Consequently, we next consider the task of symbol recovery, in which the receiver learns to recover a set of constellation points from its quantized channel output.

As shown in the previous subsection, multivariate complex-values (baseband) can be represented as real vector channels of extended dimensions. Therefore, here we focus on communications over a real-valued \ac{mimo} channel. In particular, we consider a \ac{bs} equipped with $\Nantennas = 12$ antennas, serving $\Nusers = 4$ users. The users transmit i.i.d. \ac{bpsk} symbols, represented via the vector $\myVec{s} \in \{-1,1\}^{\Nusers}$. The received signal at the \ac{bs}, denoted $\myVec{x} \in \mySet{R}^{\Nantennas}$, is given by
\begin{equation}
\label{eqn:Received1}
\myVec{x} = \myMat{H} \myVec{s} + \myVec{w},
\end{equation} 
where $\myMat{H} \in \mySet{R}^{\Nantennas \times \Nusers}$ is the channel matrix, and $\myVec{w} \in \mySet{R}^{\Nantennas}$ is additive Gaussian noise with zero-mean i.i.d. entries of variance $\sigma_w^2 > 0$.

\begin{figure}
	\centering
	\includefig{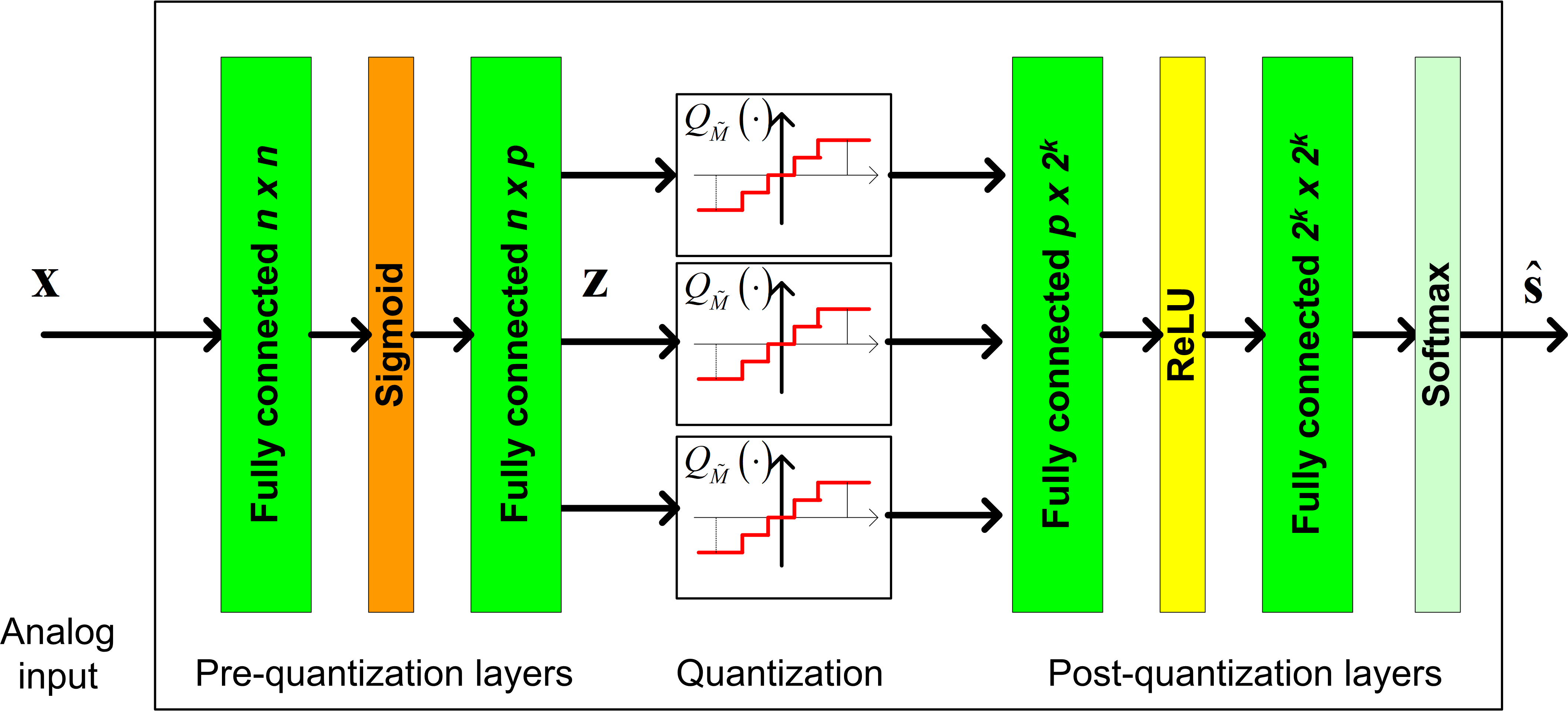} 
	\caption{Task-based symbol detector \ac{dnn}.}
	\label{fig:ClassifyNet}
\end{figure}

Here, the task of the \ac{bs} is to recover the transmitted symbols vector  $\myVec{s} $ from the channel output $\myVec{x}$, i.e., in this scenario the input dimension is $\lenX = \Nantennas$ and the task dimension is $\lenS = \Nusers$. We use a \ac{dnn} architecture consisting of two fully connected layers in analog and two fully connected layers in digital. As this is a classification task, the output layer is a softmax function with $2^{\lenS}$ probabilities, and the overall network is trained to minimize the cross-entropy loss. An illustration of the \ac{dnn} structure is depicted in Fig. \ref{fig:ClassifyNet}. Unlike the scenario considered in the previous subsection, for which the number of quantizers $\lenZ$ can be set according to the analytical results in \cite{Shlezinger:18}, here this value was determined based on empirical evaluations. In particular, we use $\lenZ = \lfloor  \lenS \Rate \rfloor$, resulting in each scalar quantizer using at least $\lenX / \lenS = 3$ bits in the hybrid system.

We compare the achievable \ac{ber} of our proposed deep task-based quantizer with soft-to-hard training to using the same architecture with passing gradient training, namely, where the quantizers are replaced with additive i.i.d. noise uniformly distributed  over the decision regions during training, which is the approach used to train neural networks with intermediate quantization in \cite{Balle:16,Balle:18,Yang:19}. In particular, for the passing gradient method we used a uniform quantization rule over the support $[-2,2]$.  The \acp{dnn} are trained using a relatively small training set consisting of $t=5000$ realizations sampled from the joint distribution of $\myVec{s}, \myVec{x}$.  

The aforementioned data-driven systems are compared to two model-based symbol detectors, which require accurate \ac{csi}, i.e.,  knowledge of $\myMat{H}$ or $\sigma_w^2$ from which the joint distribution of  $\myVec{s}$ and $\myVec{x}$ can be inferred using \eqref{eqn:Received1}: 
\begin{enumerate}
	\item The \ac{map} rule for recovering $\myVec{s}$ from $\myVec{x}$ {\em without quantization constraints}, i.e.,
	\begin{equation}
	\label{eqn:MAPrule}
	\hat{\myVec{s}}_{\rm MAP} \triangleq \mathop{\arg \max}_{\myVec{s}' \in\{-1,1\}^{\lenS}} \Pr\left(\myVec{s}=\myVec{s}' | \myVec{x}\right).
	\end{equation}
	The performance of the \ac{map} detector with perfect \ac{csi} constitutes a lower bound on the achievable \ac{ber} of any recovery scheme. 
	\item The \ac{map} rule for recovering $\myVec{s}$ from a uniformly quantized $\myVec{x}$ with rate $\Rate$, namely,
	\begin{equation}
		\label{eqn:QMAPrule}
	\hat{\myVec{s}}_{\rm QMAP} \triangleq \mathop{\arg \max}_{\myVec{s}' \in\{-1,1\}^{\lenS}} \Pr\left(\myVec{s}=\myVec{s}' | \Quan{\lfloor 2^\Rate \rfloor}{}\left( \myVec{x}\right) \right),
	\end{equation}
	where $\Quan{M'}{}(\cdot)$ represents the element-wise uniform quantization rule over the interval $[-2,2]$ using $M'$ decision regions. 
	The performance of the quantized \ac{map} detector represents the achievable \ac{ber} when processing is carried out solely in the digital domain, i.e., without using analog processing and / or tunning the quantization mapping in light of the task. 
\end{enumerate} 

Unlike the detectors based on the \ac{map} rule in \eqref{eqn:MAPrule}-\eqref{eqn:QMAPrule}, data-driven task-based quantizers do not require \ac{csi}, namely, no a-priori knowledge of $\myMat{H}$ or $\sigma_w^2$ is used in the detection procedure. Instead, a set of training samples are needed. In order to study the resiliency of our deep task-based quantizer to inaccurate training, we also compute the \ac{ber} under \ac{csi} uncertainty, namely, when the training samples are randomized from a joint distribution of $\myVec{s},\myVec{x}$ in which the entries of the matrix $\myMat{H}$ in \eqref{eqn:Received1} are corrupted by additive i.i.d. Gaussian noise, whose variance is $20\%$ the magnitude of the corresponding entry. For comparison, we also evaluate the \ac{ber} of the \ac{map} rule \eqref{eqn:MAPrule} with the same level of \ac{csi} uncertainty. The numerically computed \ac{ber} values are averaged over $20000$ Monte Carlo simulations. 

\begin{figure}
	\centering
	\includefig{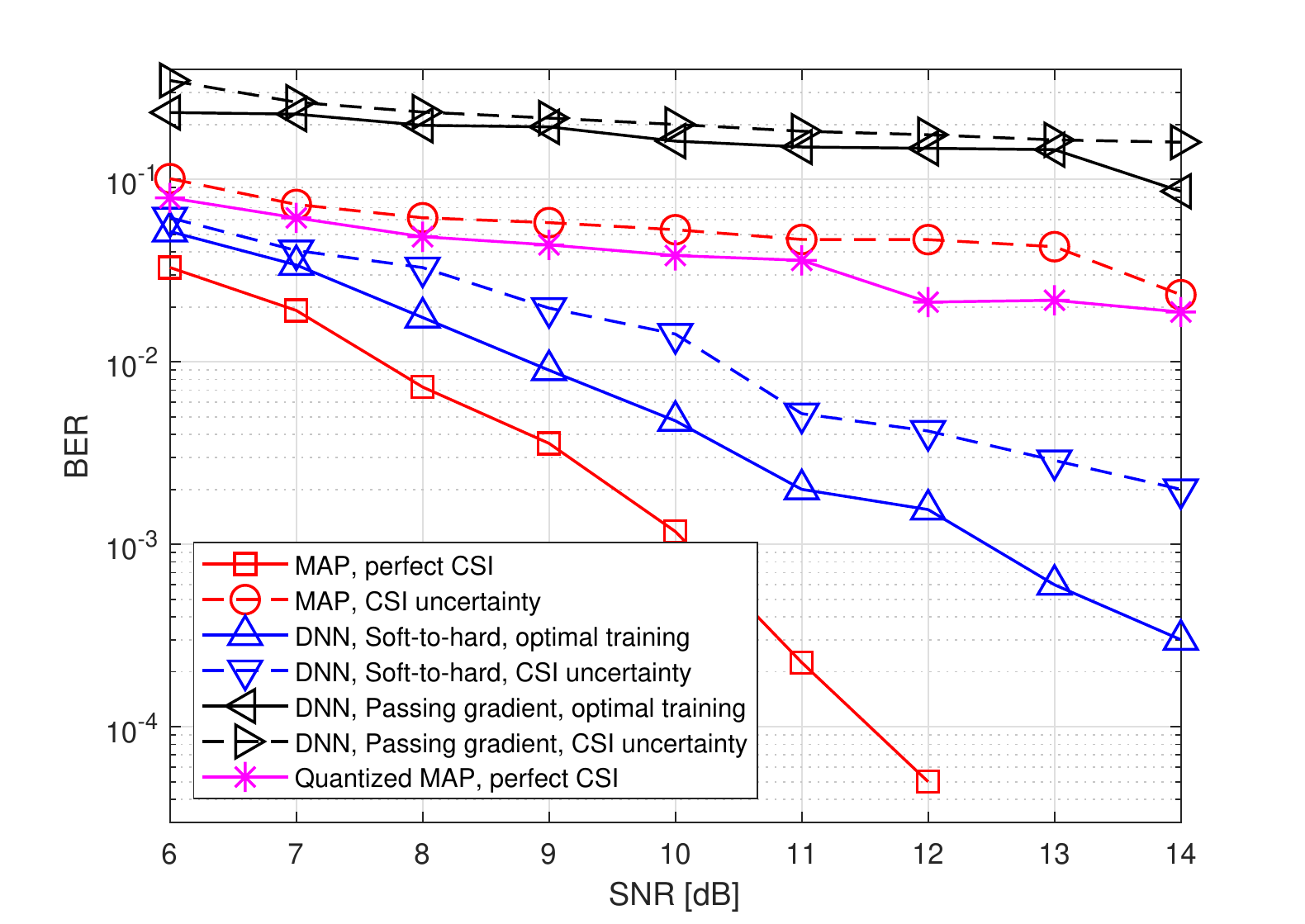} 
	\caption{\ac{ber} versus \ac{snr} at rate $\Rate = 1$.}
	\label{fig:BERvsSNR_R1}
\end{figure}

\begin{figure}
	\centering
	\includefig{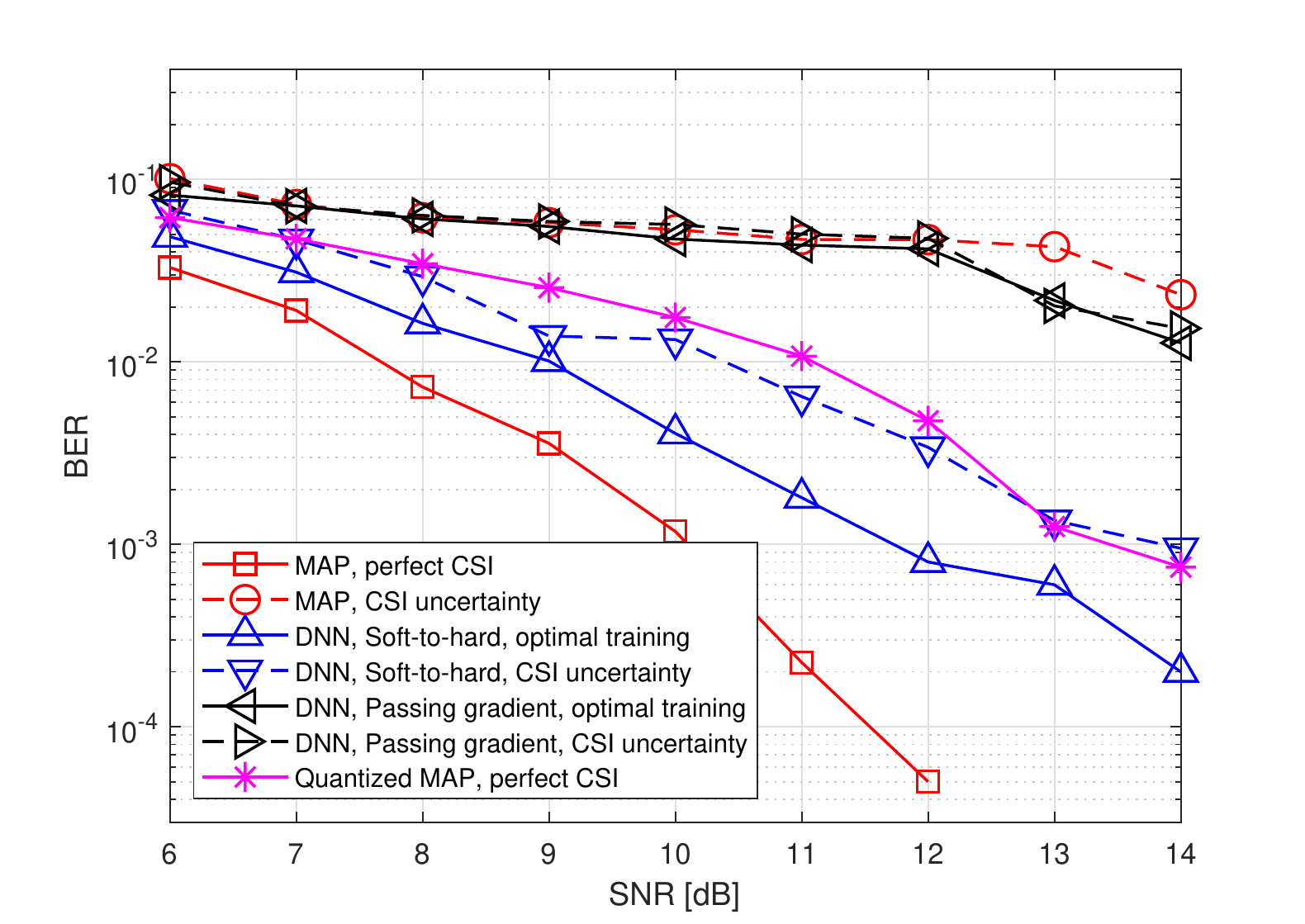} 
	\caption{\ac{ber} versus \ac{snr} at rate $\Rate = 2$.}
	\label{fig:BERvsSNR_R2}
\end{figure}

The simulated \ac{ber} values versus \ac{snr}, defined here as $1/\sigma_w^2$, in the range of $[6,14]$ dB, are depicted in Figs. \ref{fig:BERvsSNR_R1}-\ref{fig:BERvsSNR_R2} for quantization rates $\Rate =1$ and $\Rate =2$, respectively. Observing Figs. \ref{fig:BERvsSNR_R1}-\ref{fig:BERvsSNR_R2}, we note that in the presence of accurate \ac{csi}, the  \ac{ber} of our deep task-based quantizer is comparable to that achievable using the \ac{map} rule operating without quantization constraints. In particular, while the \ac{map} detector, which is independent of the quantization rate, achieves \ac{ber} of $10^{-3}$ at \ac{snr} of $10$ dB, the deep task-based quantizer obtains such \ac{ber} values at \acp{snr} of $13$ and $12$ dB, respectively, for quantization rates $\Rate =1 $ and $\Rate = 2$, respectively, namely, \ac{snr} gaps of $3$ and $2$ dB. For comparison, the quantized \ac{map} rule, which operates only in the digital domain, does not achieve \ac{ber} values below $10^{-2}$ at $\Rate=1$ and requires \ac{snr} of over $13$ dB to achieve \ac{ber} of $10^{-3}$ at rate $\Rate = 2$, i.e., with twice the number of bits used by the deep task-based quantizer to achieve the same error rate. 
This demonstrates the benefit of applying pre-quantization processing in the analog domain, which reduces the dimensionality of the input to the scalar quantizers, thus allowing to utilize more accurate quantization while keeping the semantic information required to classify the symbols from the channel output.

\begin{figure}
	\centering
	\includefig{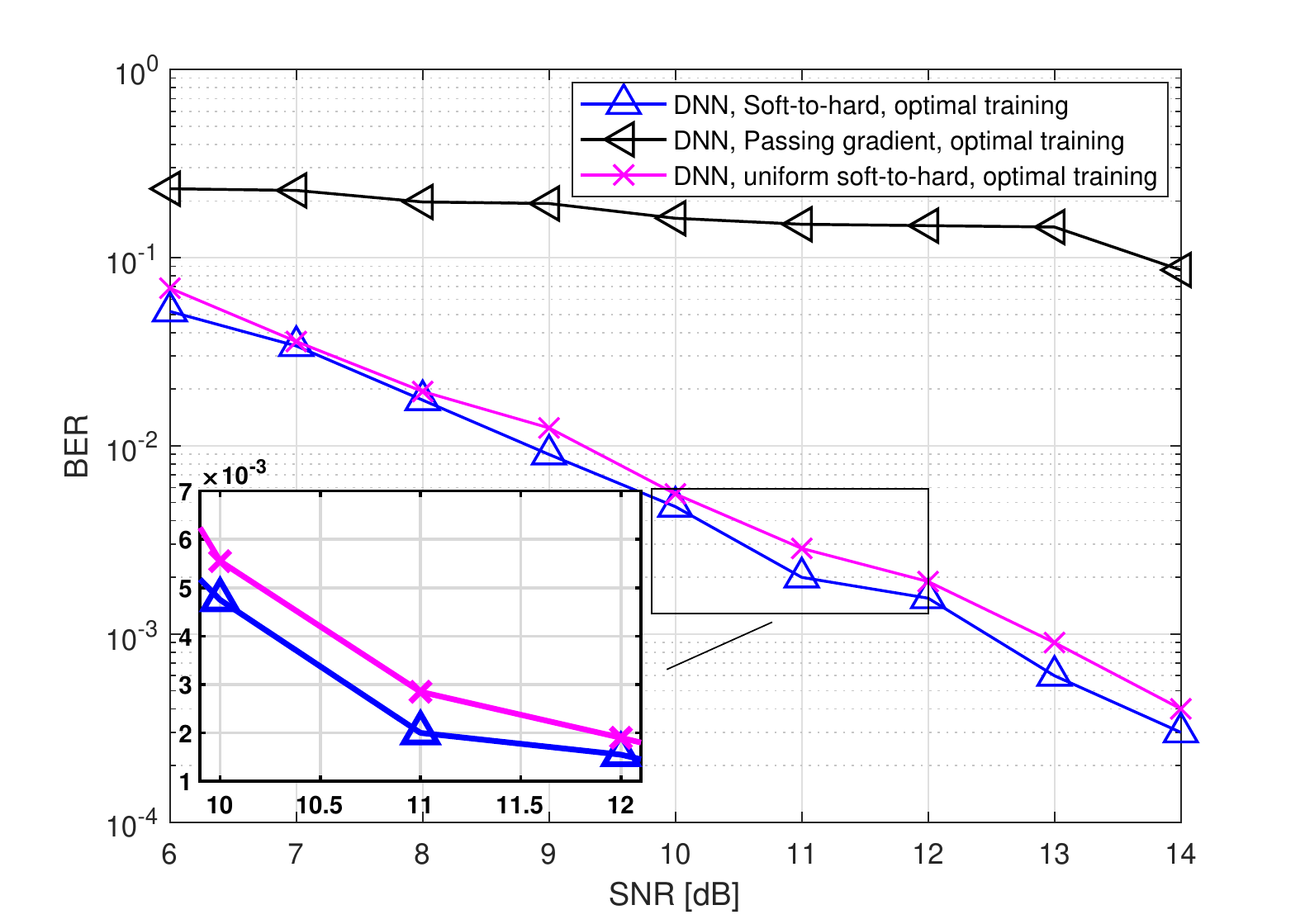} 
	\caption{\ac{ber} versus \ac{snr} at rate $\Rate = 1$}
	\label{fig:BERvsSNR_R1_CompUniform}
\end{figure}

The performance gain of the hybrid \ac{dnn} architecture stems from the ability to properly model the scalar quantizers during training using our soft-to-hard approach. This model allows to jointly train both the analog and digital \acp{dnn} as well as the decision regions of the quantizers, while accurately reflecting the quantization mapping. For comparison, it is observed in Figs. \ref{fig:BERvsSNR_R1}-\ref{fig:BERvsSNR_R2} that using the passing gradient approach, i.e., replacing quantization with additive uniformly distributed i.i.d. noise as was done in \cite{Balle:16,Balle:18,Yang:19}, leads to substantially deteriorated \ac{ber} values compared to the proposed soft-to-hard approach. 
To understand whether the improved gains of soft-to-hard modeling over passing gradient stems from the better approximation of the continuous-to-discrete mapping or from the ability to use non-uniform quantizers, we compare in Fig. \ref{fig:BERvsSNR_R1_CompUniform} the performance of the task-based quantizers with soft-to-hard modeling and with passing gradient modeling for the scenario of Fig. \ref{fig:BERvsSNR_R1} when using a fixed uniform quantizer with soft-to-hard modeling. In particular, for the uniform soft-to-hard quantizer we used the model in \eqref{eqn:tanh}  during training with the parameters $\{a_i, b_i\}$ being fixed to uniform partition of the interval $[-2,2]$, i.e., not optimized during training. It is clearly observed in Fig. \ref{fig:BERvsSNR_R1_CompUniform} that most of the gain follows from the usage of an accurate differentiable approximation of the continuous-to-discrete quantization mapping, which allows to train the system in an end-to-end manner while faithfully representing quantization. The gains due to optimizing the decision regions are rather small, indicating that our proposed approach can also lead to substantial improvements when restricted to using uniform scalar quantizers. 


The results in Figs. \ref{fig:BERvsSNR_R1}-\ref{fig:BERvsSNR_R2} also demonstrate the improved robustness to inaccurate \ac{csi}. The performance of the model-based \ac{map} detector is very sensitive to \ac{csi} uncertainty, resulting in a notable increase in \ac{ber} due to the model mismatch. However, the performance of the  deep task-based quantizer trained under  \ac{csi} uncertainty is within an \ac{snr} gap of approximately $0.5-2$ dB from its achievable performance when trained using accurate \ac{csi}. Furthermore, the deep task-based quantizer with \ac{csi} uncertainty substantially outperforms the \ac{map} rule without quantization constraints with the same level of uncertainty for all considered scenarios, and outperforms the quantized \ac{map} with accurate \ac{csi} for quantization rate of $\Rate = 1$. This demonstrates the gains of using \acp{dnn}, with their established generalization properties, for overcoming the sensitivity of model-based approaches to inaccurate knowledge of the underlying parameters.

\begin{figure}
	\centering
	\includefig{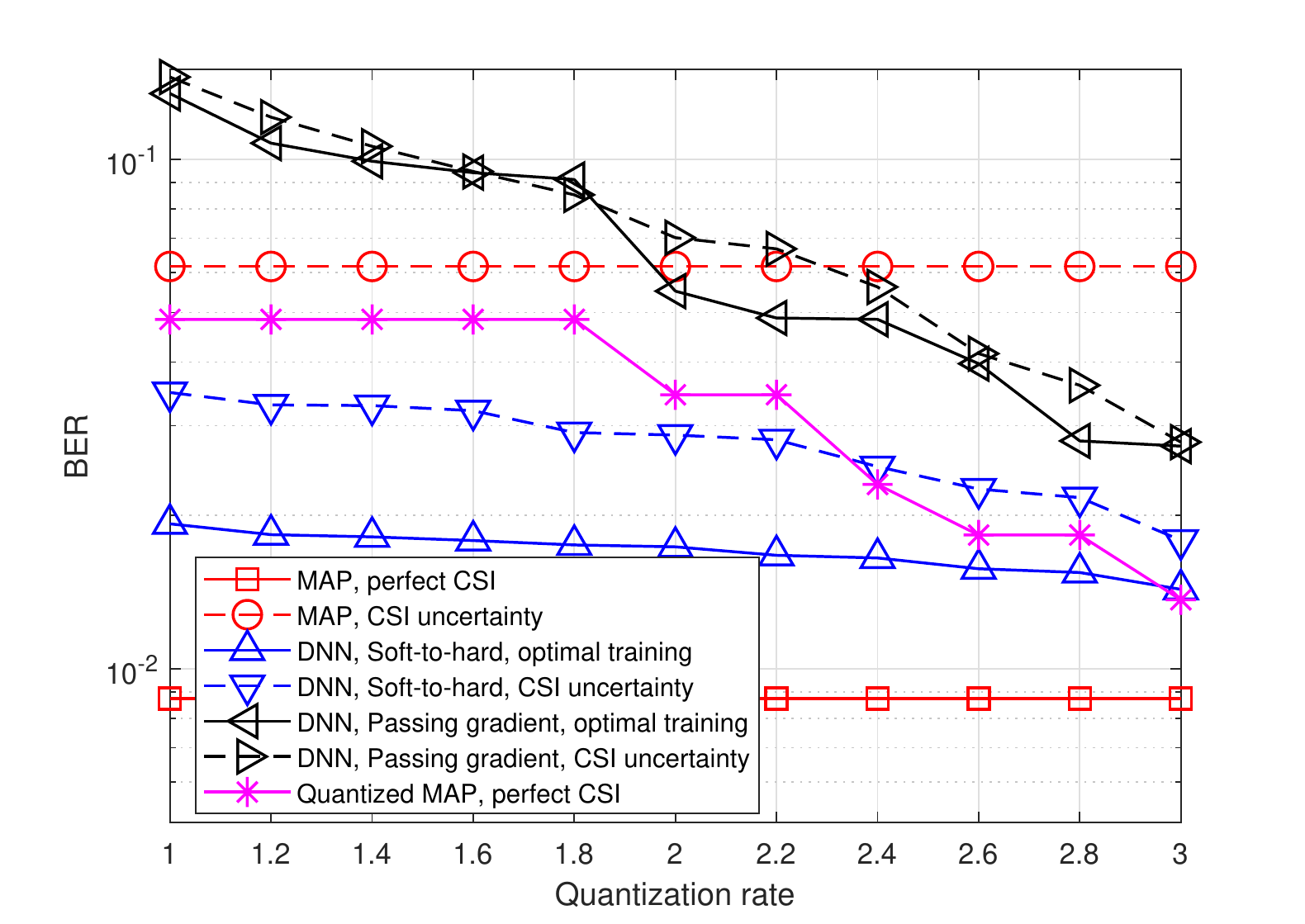} 
	\caption{\ac{ber} versus quantization rate at $8$ dB  \ac{snr}.}
	\label{fig:BERvsSNBERvsR_SNR8}
\end{figure}

\begin{figure}
	\centering
	\includefig{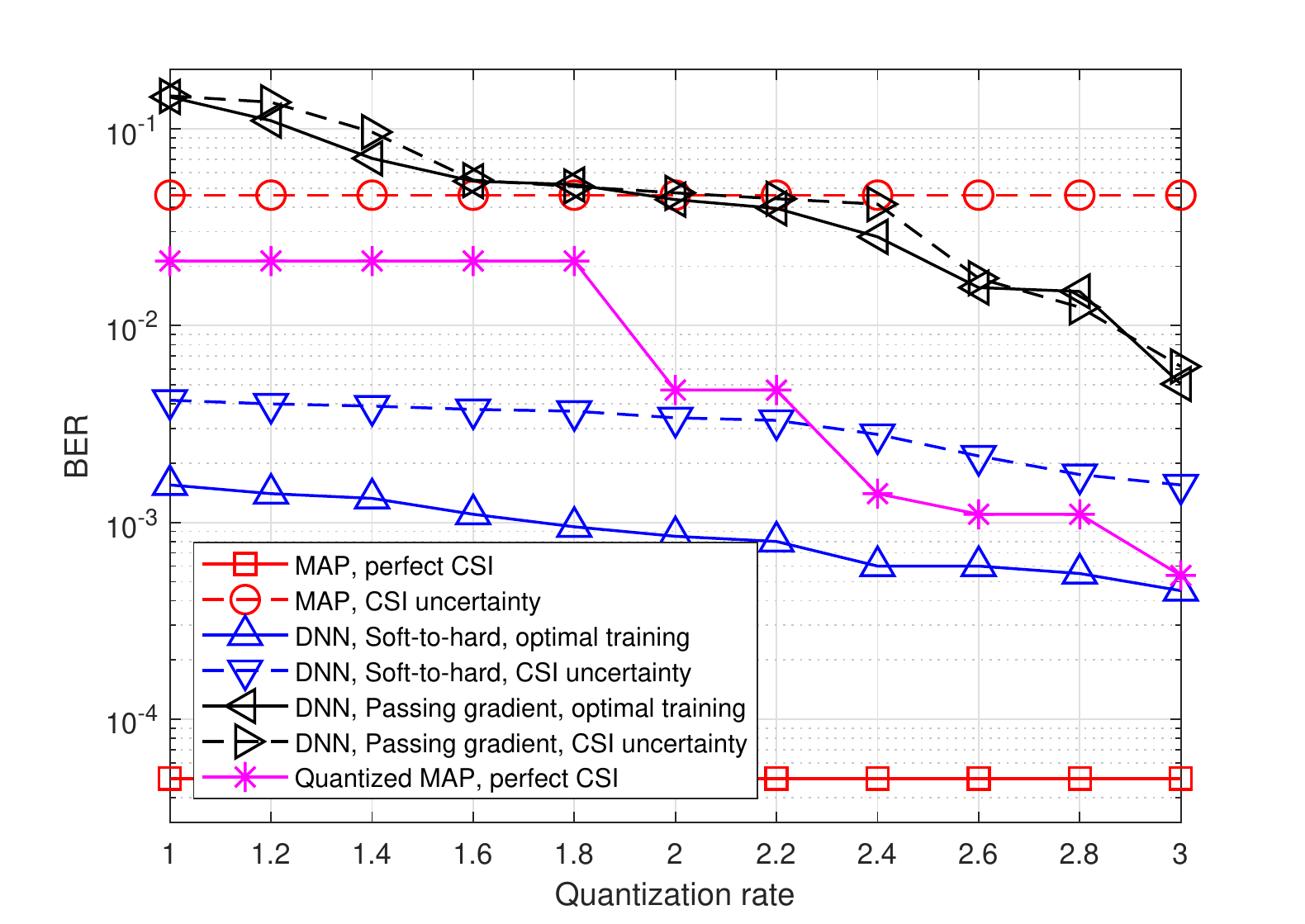} 
	\caption{\ac{ber} versus quantization rate at $12$ dB  \ac{snr}.}
	\label{fig:BERvsSNBERvsR_SNR12}
\end{figure}

Next, we evaluate the \ac{ber} of the considered quantization systems versus the quantization rate $\Rate \in [1,3]$. The results are depicted in Figs. \ref{fig:BERvsSNBERvsR_SNR8}-\ref{fig:BERvsSNBERvsR_SNR12} for \ac{snr} values of $8$ dB and $12$ dB, respectively. 
Observing Figs. \ref{fig:BERvsSNBERvsR_SNR8}-\ref{fig:BERvsSNBERvsR_SNR12}, we note that the gain of the proposed deep task-based quantizer is more dominant when operating with low quantization rates. As the quantization rate approaches three bits per channel input, the \ac{ber} of applying the \ac{map} in the digital domain via \eqref{eqn:QMAPrule} is only within a small gap of the hybrid quantizer with soft-to-hard training. However, for lower quantization rates, as well as in the presence of \ac{csi} uncertainty, the proposed deep task-based quantizer maintains its superiority observed in Figs.  \ref{fig:BERvsSNR_R1}-\ref{fig:BERvsSNR_R2}. Furthermore, it is  noted that when using the passing gradient training approach, there is a very small gap between the performance achievable with and without \ac{csi} uncertainty. This observation is likely due to the fact that when modeling quantization as additive independent noise during training, the network is trained on a mismatched model, regardless of whether the training samples are taken from the same distribution as the test samples. Consequently, such data-driven quantizers operate under some level of uncertainty even when trained using an optimal training set.

Finally, we note that the \acp{dnn} used in this subsection were trained using a relatively small training set, consisting of $t=5000$ samples. This indicates that such architectures can be used to realize an online trainable dynamic \ac{adc}, as discussed in Subsection \ref{subsec:DNNDiscussion}. 

\vspace{-0.3cm}
\section{Conclusions}
\label{sec:Conclusions}
\vspace{-0.1cm}
In this work we designed a data-driven task-based quantization system, operating with scalar \acp{adc}, using \acp{dnn}. We proposed a method for handling the non-differentiability of quantization by approximating its mapping as a smooth function. Our proposed model faithfully represents such continuous-to-discrete mappings  while allowing to learn the quantization rule from training. We discussed how this strategy can be used for designing dynamic machine-learning based \acp{adc} for various tasks. Our numerical results, which considered channel estimation and symbol recovery in bit-constrained \ac{mimo} systems, demonstrate that the performance achievable with the proposed deep task-based quantizer  is comparable with the fundamental  limits for this setup, achievable using optimal vector quantizers. Furthermore, we showed that our   soft-to-hard method for training the network in an end-to-end fashion allows the system to be accurately trained with a relatively small training set, and that it notably outperforms the common approach for training \acp{dnn} with intermediate quantization. 


\end{document}